\documentclass[a4paper,11pt]{article}
\usepackage{pos}
\usepackage{caption}
\usepackage{subcaption}
\usepackage{titlesec}

\usepackage{hyperref}
\usepackage{booktabs} 

\usepackage{tikz}
\usetikzlibrary{arrows,shapes,tikzmark,decorations.pathmorphing}
\usepackage[compat=1.1.0]{tikz-feynman}

\usepackage{diagbox}
\usepackage{adjustbox}
\newcolumntype{L}[1]{>{\raggedright\let\newline\\\arraybackslash\hspace{0pt}}m{#1}}
\newcolumntype{C}[1]{>{\centering\let\newline\\\arraybackslash\hspace{0pt}}m{#1}}
\newcolumntype{R}[1]{>{\raggedleft\let\newline\\\arraybackslash\hspace{0pt}}m{#1}}
\usepackage{multirow, bigdelim}
\usepackage{graphicx} 

\usepackage{xcolor} 
\definecolor{darkred}{rgb}{0.6, 0, 0}

\tikzset{
quark/.style={postaction={decorate}, decoration={markings}},
scalar/.style={dashed,postaction={decorate}, decoration={markings,mark=at position .5 with {\arrow[#1]{latex}}}},
gluon/.style={decorate,decoration={coil,amplitude=3pt, segment length=4.7pt, pre length=.01cm, post length=.01cm}},
gluont/.style={decorate,decoration={coil,amplitude=3pt, segment length=3.50pt, pre length=.01cm, post length=.01cm}},
}

\usepackage[utf8]{inputenc}
\DeclareUnicodeCharacter{27F9}{\ensuremath{\Longrightarrow}}
\DeclareUnicodeCharacter{2209}{\ensuremath{\notin}}
\DeclareUnicodeCharacter{266D}{\ensuremath{\flat}}
\DeclareUnicodeCharacter{2208}{\ensuremath{\in}}
\DeclareUnicodeCharacter{00E9}{\'{e}}
\DeclareUnicodeCharacter{2264}{\ensuremath{\leq}}
\DeclareUnicodeCharacter{2265}{\ensuremath{\geq}}
\DeclareUnicodeCharacter{2200}{\ensuremath{\forall}}
\DeclareUnicodeCharacter{221E}{\ensuremath{\infty}}
\DeclareUnicodeCharacter{221D}{\ensuremath{\propto}}
\DeclareUnicodeCharacter{210F}{\ensuremath{\hbar}}
\DeclareUnicodeCharacter{2020}{\ensuremath{\dagger}}
\DeclareUnicodeCharacter{03B1}{\ensuremath{\alpha}}
\DeclareUnicodeCharacter{03B2}{\ensuremath{\beta}}
\DeclareUnicodeCharacter{0393}{\ensuremath{\Gamma}}
\DeclareUnicodeCharacter{03B3}{\ensuremath{\gamma}}
\DeclareUnicodeCharacter{03B8}{\ensuremath{\theta}}
\DeclareUnicodeCharacter{21D2}{\ensuremath{\Rightarrow}}
\DeclareUnicodeCharacter{0394}{\ensuremath{\Delta}}
\DeclareUnicodeCharacter{03B4}{\ensuremath{\delta}}
\DeclareUnicodeCharacter{03C1}{\ensuremath{\rho}}
\DeclareUnicodeCharacter{03BE}{\ensuremath{\xi}}
\DeclareUnicodeCharacter{03C0}{\ensuremath{\pi}}
\DeclareUnicodeCharacter{03A0}{\ensuremath{\Pi}}
\DeclareUnicodeCharacter{03BC}{\ensuremath{\mu}}
\DeclareUnicodeCharacter{03BD}{\ensuremath{\nu}}
\DeclareUnicodeCharacter{222B}{\ensuremath{\int}}
\DeclareUnicodeCharacter{2211}{\ensuremath{\Sigma}}
\DeclareUnicodeCharacter{03C3}{\ensuremath{\sigma}}
\DeclareUnicodeCharacter{03C4}{\ensuremath{\tau}}
\DeclareUnicodeCharacter{03BB}{\ensuremath{\lambda}}
\DeclareUnicodeCharacter{03B7}{\ensuremath{\eta}}
\DeclareUnicodeCharacter{03A6}{\ensuremath{\Phi}}
\DeclareUnicodeCharacter{03D5}{\ensuremath{\phi}}
\DeclareUnicodeCharacter{03A8}{\ensuremath{\Psi}}
\DeclareUnicodeCharacter{03C8}{\ensuremath{\psi}}
\DeclareUnicodeCharacter{03F5}{\ensuremath{\epsilon}}
\DeclareUnicodeCharacter{2202}{\ensuremath{\partial}}
\DeclareUnicodeCharacter{220F}{\ensuremath{\prod}}
\DeclareUnicodeCharacter{03C9}{\ensuremath{\omega}}
\DeclareUnicodeCharacter{2260}{\ensuremath{\neq}}
\DeclareUnicodeCharacter{22C5}{\ensuremath{\cdot}}
\DeclareUnicodeCharacter{27E8}{\ensuremath{\langle}}
\DeclareUnicodeCharacter{27E9}{\ensuremath{\rangle}}
\DeclareUnicodeCharacter{2192}{\ensuremath{\rightarrow}}
\DeclareUnicodeCharacter{00B2}{\ensuremath{{}^2}}
\DeclareUnicodeCharacter{221A}{\ensuremath{\sqrt{}}}
\DeclareUnicodeCharacter{223C}{\ensuremath{\sim}}
\DeclareUnicodeCharacter{226A}{\ensuremath{\ll}}
\DeclareUnicodeCharacter{00D7}{\ensuremath{\times}}

\title{Non-Planar Two-Loop Amplitudes \\ for Five-Parton Scattering}
\ShortTitle{Non-Planar Two-Loop Amplitudes for Five-Parton Scattering}

\author[a]{Giuseppe De Laurentis}

\affiliation[a]{Higgs Centre for Theoretical Physics, University of Edinburgh, \\
  Edinburgh, EH9 3FD, United Kingdom
}

\emailAdd{giuseppe.delaurentis@ed.ac.uk}

\abstract{
  We review the current status of high-multiplicity double-virtual QCD
  corrections to processes relevant for LHC phenomenology. In
  particular, we discuss the recent full-color calculation of the
  five-parton process, whose two-loop amplitudes are required to
  obtain next-to-next-to-leading order predictions for three-jet
  production at the LHC. We address various aspects of the
  computation, including color decomposition, renormalization, partial
  amplitudes, color identities and the construction of the finite
  remainder. We review the method of numerical unitarity, which is
  used to generate finite-field samples of the amplitude. We then
  focus on the analytic reconstruction of the coefficient functions
  from these numerical samples via Ansatz techniques. A novel
  algorithm, based on the correlation of codimension-one residues,
  helps manage the complexity of the calculation. Little-group
  rescalings of the gluon amplitude, inspired by supersymmetry Ward
  identities, facilitate the computation of the quark amplitudes. We
  conclude with an outlook towards upcoming computations with an
  increased number of scales, leading to larger Ans\"atze and more
  complicated alphabets.
  }

\FullConference{%
  Loops and Legs in Quantum Field Theory - LL2024,\\
  14-19 April, 2024\\
  Wittenberg, Germany
}

\begin{document}
\maketitle

\section{Introduction}
In the absence of glaring signals of new physics (NP) at the Large
Hadron Collider (LHC), precision studies of the Standard Model (SM) of
particle physics have been playing an increasingly important role. The
aim is to better understand aspects of the SM, such as the convergence
of the first orders in the perturbative expansion, factorization
properties, and values of the free parameters, and at the same time
increase the sensitivity to subtle deviations that may hint at
NP. Precise theoretical predictions play a crucial role in both
regards. In particular, next-to-next-to-leading order (NNLO)
predictions in quantum chromodynamics (QCD) are essential for
precision studies targeting $\sim\kern-1mm1\%$ uncertainties, and in
some cases, N$^3$LO predictions may even be desirable
\cite{Begel:2022kwp, Andersen:2024czj}.

Recent years have seen significant advancements in the computation of
higher-order QCD corrections for multi-scale processes. These consist
predominantly of two-loop five-point amplitudes, with massless
propagators and up to one external massive leg. These amplitudes were
first obtained in the leading color (l.c.) approximation. For pure
QCD processes, this generally corresponds to planar diagrams. However,
especially in the presence of electroweak particles, non-planar
diagrams may contribute at leading color. Nevertheless, it is usually
possible to identify gauge-invariant planar subamplitudes. As of the
end of 2023, all five-point massless processes relevant for
hadron-collider phenomenology have been obtained in full color, while
no five-point one-mass process is known beyond the planar
approximation. Table \ref{tab:five-point-summary} summarizes the
current status.

\vspace{0mm}
\begin{table}[h!]
    \centering
    \begin{tabular}{p{8em} l l l}
        \toprule
        \textbf{Processes} & \textbf{Analytic Results} & \textbf{Public Codes} & \textbf{Cross Sections} \\
        \midrule

        $pp \rightarrow \gamma\gamma\gamma$ & \cite{Abreu:2020cwb,
          Chawdhry:2020for}$^{\ddagger}$ \bf\cite{Abreu:2023bdp} &
        \cite{Abreu:2020cwb}$^{\ddagger}$, \bf\cite{Abreu:2023bdp} &
        \cite{Chawdhry:2019bji, Kallweit:2020gcp}$^{\ddagger}$ \\
        
        $pp \rightarrow \gamma\gamma j$ & \cite{Agarwal:2021grm,
          Chawdhry:2021mkw,Badger:2021imn}$^{\dagger}$
        \bf\cite{Agarwal:2021vdh} & \cite{Agarwal:2021grm,Badger:2021imn}$^{\dagger}$
        & \cite{Chawdhry:2021hkp, Badger:2021ohm}$^{\dagger}$ \\
        
        $pp \rightarrow \gamma jj$ & \bf\cite{Badger:2023mgf} & &
        \bf\cite{Badger:2023mgf} \\

        $pp \rightarrow jjj$ &
        \cite{Abreu:2021oya}$^{\dagger}$\bf\cite{Agarwal:2023suw,
          DeLaurentis:2023nss, DeLaurentis:2023izi} &
        \cite{Abreu:2021oya}$^{\dagger}$\bf\cite{DeLaurentis:2023izi}
        & \cite{Czakon:2021mjy, Chen:2022ktf}$^{\dagger}$ \\

        \midrule $pp \rightarrow Wb\bar{b}$ &
        \cite{Badger:2021nhg}$^{\star}$ \cite{Abreu:2021asb,
          Hartanto:2022qhh,DeLaurentis:2024xxx}$^{\dagger}$ &
        \cite{DeLaurentis:2024xxx}$^\dagger$ & \cite{Hartanto:2022qhh,
          Hartanto:2022ypo}$^{\dagger}$ \\

        $pp \rightarrow Hb\bar{b}$ & \cite{Badger:2021ega}$^{\star}$ & & \\

        $pp \rightarrow Wj\gamma$ & \cite{Badger:2022ncb}$^{\ddagger}$ & & \\

        $pp \rightarrow Wjj$ &
        \cite{Abreu:2021asb,DeLaurentis:2024xxx}$^{\dagger}$ &
        \cite{DeLaurentis:2024xxx}$^\dagger$ & \\
        
        $pp \rightarrow (Z/\gamma^\star)jj$ &
        \cite{Abreu:2021asb,DeLaurentis:2024xxx}$^{\ddagger}$ &
        \cite{DeLaurentis:2024xxx}$^\ddagger$ & \\

        \midrule

        $pp \rightarrow ttH$ & & & \cite{Catani:2022mfv}$^{\star}$ \\

        \bottomrule
    \end{tabular}
    \caption{
      Summary of known two-loop QCD corrections for five-point
      scattering processes at hadron colliders. \linebreak ${\dagger}$
      denotes calculations performed in l.c.~approximation, where
      l.c.~coincides with planar, while ${\ddagger}$ denotes planar
      computations that are not l.c.~accurate. ${\star}$ denotes
      additional approximations, such as on-shell $W$, $m_b = 0$ but
      $y_b \neq 0$, or soft Higgs. Bold denotes non-planar, full-color
      results.
    }
    \label{tab:five-point-summary}
    \vspace{-2.5mm}
\end{table}

The computation of these amplitudes was made possible thanks to
developments on two main fronts. First, regarding the transcendental
part of loop amplitudes, this includes the application of differential
equations to Feynman integrals
\cite{KOTIKOV1991123,KOTIKOV1991158,Remiddi:1997ny}, their
$\epsilon$-factorized form \cite{Henn:2013pwa}, and progress on
integration-by-parts (IBP) reduction \cite{CHETYRKIN1981159} using
algebro-geometric techniques
\cite{Gluza:2010ws,Ita:2015tya,Larsen:2015ped}. To enable
phenomenological applications, the development of numerically
efficient and stable implementations of master integrals was
crucial. This was achieved through Chen's iterated integrals
\cite{Chen:1977oja}, implemented in so-called pentagon functions
\cite{Gehrmann:2018yef} now available in the public code
\texttt{PentagonFunctions++} for both massless
\cite{Chicherin:2020oor} and one-mass \cite{Chicherin:2021dyp,
  Abreu:2023rco} five-point processes.

Second, on the rational coefficient side, a crucial development was
the introduction of exact numerical arithmetic, in the form of finite
fields \cite{vonManteuffel:2014ixa, Peraro:2016wsq}, and
reconstruction techniques based on interpolation \cite{Peraro:2016wsq,
  Peraro:2019svx} or Ansatz fitting \cite{DeLaurentis:2019bjh,
  DeLaurentis:2022otd}. To achieve compact results, it is important to
control the evaluation of the coefficients in degenerate kinematic
limits in complex kinematics \cite{DeLaurentis:2019bjh}, which
generalize the familiar concepts of soft and collinear limits. To
reconcile this type of evaluations with exact arithmetic the use of
$p$-adic numbers was proposed \cite{DeLaurentis:2022otd} and then
further investigated in relation to multivariate partial fraction
decompositions \cite{Campbell:2022qpq, Chawdhry:2023yyx}. Another
recent development was the use of $\mathbb{Q}[\vec x]$ linear
relations \cite{Liu:2023cgs} to facilitate reconstruction. In this
work, we made use of a similar concept, namely $\mathbb{Q}$-linear
relations valid in singular limits.
\vspace{-1mm}

\section{Scattering Amplitudes and Finite Remainders}
\vspace{-1mm}

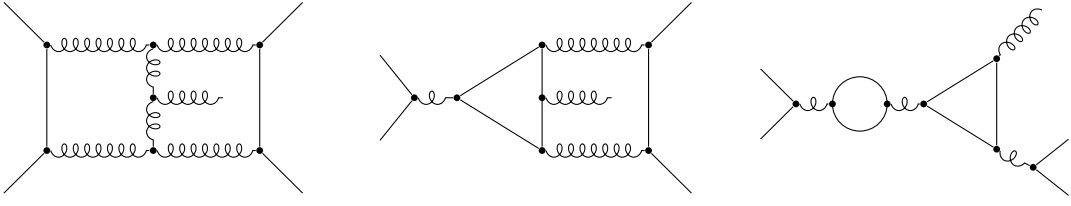
\begin{figure}[t]
\centering \begin{subfigure}{0.31\linewidth}
  \begin{tikzpicture}[scale=0.7]
    \begin{feynman}[medium]
      \vertex (g1) at (-1,1) {};
      \vertex (g2) at (-1,-3) {};
      \vertex (g3) at (5,1) {};
      \vertex (g4) at (5,-3) {};
      \vertex (g5) at (3.5,-1) {};
      \vertex (a1)[circle, fill, inner sep=0.9pt] at (0,0) {};
      \vertex (a2)[circle, fill, inner sep=0.9pt] at (0,-2) {};
      \vertex (b1)[circle, fill, inner sep=0.9pt] at (2,0) {};
      \vertex (b3)[circle, fill, inner sep=0.9pt] at (2,-1) {};
      \vertex (b2)[circle, fill, inner sep=0.9pt] at (2,-2) {};
      \vertex (c1)[circle, fill, inner sep=0.9pt] at (4,0) {};
      \vertex (c2)[circle, fill, inner sep=0.9pt] at (4,-2) {};
      \diagram* {
        (g1) -- [quark] (a1);
        (c1) -- [quark] (g3);
        (a1) -- [gluon] (b1) -- [gluon] (c1);
        (g2) -- [quark] (a2);
        (c2) -- [quark] (g4);
        (a2) -- [gluon] (b2) -- [gluon] (c2);
        (a1) -- [quark] (a2);
        (b1) -- [gluon] (b3) -- [gluon] (b2);
        (b3) -- [gluon] (g5);
        (c1) -- [quark] (c2);
      };
    \end{feynman} 
  \end{tikzpicture}
\end{subfigure}
~
\begin{subfigure}{0.31\linewidth}
  \begin{tikzpicture}[scale=0.7]
    \begin{feynman}[medium]
    \vertex (g1) at (-1.2,0) {};
    \vertex (g2) at (-1.2,-2) {};
    \vertex (g3) at (5,1) {};
    \vertex (g4) at (5,-3) {};
    \vertex (g5) at (3.5,-1) {};
    \vertex (a1)[circle, fill, inner sep=0.9pt] at (-0.4,-1) {};
    \vertex (a2)[circle, fill, inner sep=0.9pt] at (0.4,-1) {};
    \vertex (b1)[circle, fill, inner sep=0.9pt] at (2,0) {};
    \vertex (b3)[circle, fill, inner sep=0.9pt] at (2,-1) {};
    \vertex (b2)[circle, fill, inner sep=0.9pt] at (2,-2) {};
    \vertex (c1)[circle, fill, inner sep=0.9pt] at (4,0) {};
    \vertex (c2)[circle, fill, inner sep=0.9pt] at (4,-2) {};
    \diagram* {
      (g1) -- [quark] (a1) -- [quark](g2);
      (g3) -- [quark] (c1) -- [quark] (c2) -- [quark] (g4);
      (a1) -- [gluon] (a2);
      (a2) -- [quark] (b1) -- [quark] (b2) -- [quark] (a2);
      (b2) -- [gluon] (c2);
      (b1) -- [gluon] (c1);
      (b3) -- [gluon] (g5);
    };
    \end{feynman} 
  \end{tikzpicture}
\end{subfigure}
~
\begin{subfigure}{0.31\linewidth}
  \begin{tikzpicture}[scale=0.6]
    \begin{feynman}[medium]
      \vertex (g1) at (-1.4,0) {};
      \vertex (g2) at (-1.4,-2) {};
      \vertex (g3) at (5.2,1.3) {};
      \vertex (g4) at (5.8,-1.6) {};
      \vertex (g5) at (5.8,-3.2) {};
      \vertex (a1)[circle, fill, inner sep=0.9pt] at (-0.4,-1) {};
      \vertex (a2)[circle, fill, inner sep=0.9pt] at (0.4,-1) {};
      \vertex (b1)[circle, fill, inner sep=0.9pt] at (1.6,-1) {};
      \vertex (b2)[circle, fill, inner sep=0.9pt] at (2.4,-1) {};
      \vertex (c1)[circle, fill, inner sep=0.9pt] at (4,0) {};
      \vertex (c2)[circle, fill, inner sep=0.9pt] at (4,-2) {};
      \vertex (d1)[circle, fill, inner sep=0.9pt] at (4.8,-2.4) {};
      \draw [quark] (b1) arc (0:360:0.6);
      \diagram* {
        (a1) -- [gluon] (a2);
        (b1) -- [gluon] (b2);
        (g1) -- [quark] (a1);
        (g2) -- [quark] (a1);
        (g3) -- [gluon] (c1);
        (g4) -- [quark] (d1) -- [quark] (g5);
        (c2) -- [gluon] (d1);
        (b2) -- [quark] (c1) -- [quark] (c2) -- [quark] (b2); 
      };
    \end{feynman} 
  \end{tikzpicture}
\end{subfigure}
\caption{Representative Feynman diagrams for two-loop four-quark
  one-gluon amplitudes, contributing at $N_f^0$, $N_f^1$ and $N_f^2$
  respectively. Solid lines represent massless quarks.}
\label{tab:diagram_4q1g}
\end{figure}

The two-loop amplitudes required for NNLO predictions of
$pp\rightarrow jjj$ in full color were recently computed in
\cite{Agarwal:2023suw, DeLaurentis:2023nss, DeLaurentis:2023izi}.  We
follow in detail the approach of \cite{DeLaurentis:2023nss,
  DeLaurentis:2023izi}. Three partonic processes are required,
\vspace{-1.9mm}
\begin{gather}
 g_{-p_1}^{-h_1}+g_{-p_2}^{-h_2} \to g_{p_3}^{h_3}+g_{p_4}^{h_4}+g_{p_5}^{h_5} \,, \label{eq:5g} \\
 \bar{u}_{-p_1}^{-h_1}+u_{-p_2}^{-h_2} \to g_{p_3}^{h_3} +g_{p_4}^{h_4}+g_{p_5}^{h_5} \,, \label{eq:2q3g} \\
 \bar{u}_{-p_1}^{-h_1}+u_{-p_2}^{-h_2} \to d_{p_3}^{h_3} +\bar{d}_{p_4}^{h_4}+g_{p_5}^{h_5} \,, \label{eq:4q1g}
\end{gather}
including all crossings thereof, obtained via analytic continuation,
as well as the process with four identical quark flavors, obtained
from linear combinations of eq.~\eqref{eq:4q1g}. As indicated by the
labels for momenta ($p_i$) and helicities ($h_i$), we work in the
all-outgoing convention.

\vspace{-0.8mm}
\paragraph{Color decomposition}
The amplitude for each partonic process admits a color decomposition
in terms of fundamental $SU(N_c)$ generators. For the four-quark
one-gluon process, it reads,
\begin{align}\label{eq:partial_4q}
  \mathcal{A}_{\vec{a}}(1_u,2_{\bar u},3_d,4_{\bar d},5_g) = & 
 \sum_{\sigma \in \mathcal{Z}_2(\{1,2\},\{3,4\})} \sigma\Big(\delta^{\bar i_4}_{i_1} (T^{a_5})^{\,\bar i_2}_{i_3} \; A_{d}(1,2,3,4,5)\Big) \; + \nonumber \\
 & \sum_{\sigma \in \mathcal{Z}_2(\{1,2\},\{3,4\})} \kern-2mm \sigma\Big( \delta^{\bar i_2}_{i_1} (T^{a_5})^{\,\bar i_4}_{i_3} \; A_{s}(1,2,3,4,5)\Big) \, ,
\end{align}
where the color-ordered sub-amplitudes $A_d$ and $A_s$ are labelled
depending on whether the color-space Kronecker delta is between
different or the same quark line, respectively. Permutations are
denoted as $\sigma$, and $\mathcal{Z}_n$ is the cyclic group of order
$n$.

\paragraph{Renormalization}
Each color-ordered amplitude further admits an asymptotic expansion in
powers of the bare QCD coupling $\alpha^0_s$. Representative diagrams
with a maximum number of propagators for the four-quark one-gluon
process at order $(\alpha^0_s)^2$ are displayed in
fig.~\ref{tab:diagram_4q1g}. We perform renormalization in the
$\overline{\text{MS}}$ scheme, via the substitution,
\begin{equation}\label{eq:renorm}
  \alpha^0_s\mu_0^{2\epsilon}S_{\epsilon}
  =\alpha_s\mu^{2\epsilon}
  \left(
  1-\frac{\beta_0}{2\epsilon}\frac{\alpha_s}{2\pi}
  +\left(\frac{\beta_0^2}{4 \epsilon^2}-\frac{\beta_1}{ 8 \epsilon}\right) \left(\frac{\alpha_s}{2\pi}\right)^2+\mathcal{O}\left(\alpha_s^3\right)\right)\,, \nonumber
\end{equation}

\paragraph{Partial Amplitudes}
At each order in the $\alpha^0_s$ (or $\alpha_s$) expansion,
amplitudes can be further expanded in powers of the number of colors,
$N_c^{n_c}$, and of the number of light quark flavors, $N_f^{n_f}$. We
label these gauge-invariant building blocks with the notation
$A^{(L),(n_c, n_f)}$. For the two four-quark one-gluon amplitudes of
eq.~\ref{eq:partial_4q} at two-loop this reads,
\begin{gather}
  \scalebox{0.9}{$\displaystyle A_d^{(2)} = N_c^2 A_d^{(2),(2,0)} + {\color{darkred} A_d^{(2),(0,0)}} + \frac{1}{N_c^2} {\color{darkred} A_d^{(2),(-2,0)}}
  +  N_f N_c A_d^{(2),(1,1)} + \frac{N_f}{N_c} {\color{darkred} A_d^{(2),(-1,1)}} + N_f^2  A_d^{(2),(0,2)} \, ,$} \\
  \scalebox{0.9}{$\displaystyle A_s^{(2)} = N_c {\color{darkred} A_s^{(2),(1,0)}}+\frac{1}{N_c}{\color{darkred} A_s^{(2),(-1,0)}}+\frac{1}{N_c^3}{\color{darkred} A_s^{(2),(-3,0)}}
  + N_f{\color{darkred} A_s^{(2),(0,1)}} + \frac{N_f}{N_c^2} {\color{darkred} A_s^{(2),(-2,1)}} + \frac{N_f^2}{N_c}{\color{darkred} A_s^{(2),(-1,2)}} \, .$}
\end{gather}
The subleading-color partial amplitudes displayed in red were obtained
recently \cite{Agarwal:2023suw, DeLaurentis:2023izi}.

\paragraph{Color identities}
It is well known that the color decomposition of
eq.~\ref{eq:partial_4q} is not optimal in terms of having independent
partial amplitudes. This is not an issue for our computation strategy,
since its cost is largely unaffected by redundancies among
partials. Nevertheless, we were interested in uncovering potentially
new relations, such as
\begin{gather}
  \big[ 32 \, A^{(2),(2,0)}_d \, (1,2,3,4,5) + 8 \, A^{(2),(0,0)}_d \, (1,2,3,4,5) + 2 A^{(2),(-2,0)}_d \, (1,2,3,4,5) \\
    + 16 \, A^{(2),(1,0)}_s \, (1,2,3,4,5) \, + 4 A^{(2),(-1,0)}_s \, ( 1,2,3,4,5)
    + A^{(2),(-3,0)}_s  \, (1,2,3,4,5) \big] - \big[ \dots \big]_{3
    \leftrightarrow 4}= 0 \, . \nonumber
\end{gather}
We obtain such relations via numerical finite-field evaluations and
linear algebra (null-space computation and intersection). Note that
while an identity exists among partials, it does not allow to express
one amplitude as a linear combination of others, unlike in the pure
Yang-Mills channel. This is due to the fact that each partial
amplitude appears with multiple permutations of the arguments.

\paragraph{Finite remainder}
We subtract the infrared singularities via a color-space operator
${\bf Z}(\epsilon, \mu)$ \cite{Becher:2009cu} (see also
\cite{Catani:1998bh, Gardi:2009qi}),
\begin{equation}\label{eqn:remainder}
  {\mathcal R}(\lambda, \tilde\lambda, \mu) ={\bf Z}(\epsilon, \lambda\tilde\lambda, \mu) {\mathcal A}(\lambda, \tilde\lambda, \mu) ~+~ \mathcal{O}({\epsilon}) \,.
\end{equation}
$\mathcal R$ denotes the so-called finite remainder, i.e.~it is the
finite part of the amplitude carrying the new physical information at
a given loop order. It can be written as a weighted sum of pentagon
functions $h_i$ with rational function coefficients $r_i$,
\begin{equation}
  \mathcal{R}(\lambda, \tilde\lambda, \mu) = \sum_i r_{i}(\lambda,\tilde\lambda) \, h_i(\lambda\tilde\lambda, \mu) \, .
\end{equation}
Note that knowledge of the one-loop amplitude $\mathcal{A}^{(1)}$ at
higher orders in the dimensional regulator $\epsilon$ is only required
to build the two-loop finite remainder $\mathcal{R}^{(2)}$. For the
computation of NNLO observables, such as differential cross-sections,
the finite remainders $\mathcal{R}^{(1)}$ and $\mathcal{R}^{(2)}$
suffice \cite{Weinzierl:2011uz} (besides the tree
$\mathcal{A}^{(0)}$).

\section{Numerical Computation over Finite Fields}

To circumvent the swell of complexity in intermediate stages of the
computation, we first compute partial amplitudes and associated finite
remainders numerically over finite fields. This is achieved via the
method of numerical unitarity
\cite{Ita:2015tya,Abreu:2017idw,Abreu:2017xsl,Abreu:2018jgq}, as
implemented in the public code {\sc Caravel} \cite{Abreu:2020xvt}. The
amplitude integrand $A(\lambda, \tilde\lambda, \ell)$ is expressed as
a linear combination of topologies $\Gamma$ and, for each topology, a
set of master integrands $M_\Gamma$ and surface terms $S_\Gamma$, with
coefficients $c_{\,\Gamma,i}(\lambda, \tilde\lambda, \epsilon)$,
dependent on external kinematics and epsilon, namely,
\begin{equation}
  A(\lambda, \tilde\lambda, \ell) \quad = \quad\; \sum_{\substack{\Gamma, \, i \in M_\Gamma \cup S_\Gamma}} \, c_{\,\Gamma,i}(\lambda, \tilde\lambda, \epsilon) \,		\frac{m_{\Gamma,i}(\lambda\tilde\lambda, \ell)}{\textstyle \prod_{j} \rho_{\,\Gamma,j}(\lambda\tilde\lambda, \ell)} \, .
\end{equation}

After integration, surface terms drop out (they integrate to zero),
master integrands yield master integrals $I_{\Gamma, i}$, and, for a
suitable choice of these, the location of the poles in the dimensional
regulator $\epsilon$ take values independent of the external
kinematics ($a_{ij} \in \mathbb{Q}$),
\begin{equation}
  A(\lambda, \tilde\lambda) = \int d^D\ell \; A(\lambda, \tilde\lambda, \ell) = \sum_{\substack{\Gamma,\\ i \in M_\Gamma}} \frac{ \sum_{k=0}^{k_{\text{max}}} \, {\color{black}c^{(k)}_{\,\Gamma, i}}(\lambda, \tilde\lambda) \, \epsilon^k}{\prod_j (\epsilon - a_{ij})} \, {\color{black}I_{\Gamma, i}}(\lambda\tilde\lambda, \epsilon)
\end{equation}

In a procedure equivalent to IBP reduction, the coefficients of the
master integrands (and surface terms) are obtained by solving linear
systems of equations to match the right-hand side parametrization of
the integrand to products of trees on cuts,
\begin{equation}
  \displaystyle \sum_{\text{states}} \, \prod_{\text{trees}} A^{\text{tree}}(\lambda, \tilde\lambda, \ell)\big|_{\Gamma_{\text{cut}}} = \sum_{\substack{\Gamma' \ge \Gamma_{\text{cut}}, \\ i \in M_\Gamma' \cup S_\Gamma'}} \kern-2mm c_{\,\Gamma',i}(\lambda, \tilde\lambda) \, \frac{m_{\Gamma',i}(\lambda\tilde\lambda, \ell)}{\displaystyle \prod_{j\in P_{\Gamma'} / P_{\Gamma_{\text{cut}}}} \rho_{j}(\lambda\tilde\lambda, \ell)}\Bigg|_{\Gamma_{\text{cut}}} \, .
\end{equation}
This requires prior knowledge of the integrand decomposition $M_\Gamma
\cup S_\Gamma$. For the computation of $pp\rightarrow jjj$ at two-loop
this decomposition was obtained by extending to higher power counting
the decomposition previously obtained for $pp\rightarrow
\gamma\gamma\gamma$ using the embedding space formalism
\cite{Abreu:2023bdp}.

\section{Analytic Reconstruction}

To enable phenomenological studies, we must have access to reasonably
fast and stable evaluations of the amplitude (or finite remainder)
with floating-point numbers. This is achieved by reconstructing
compact analytic form from numerical evaluations in $\mathbb{F}_p$
\cite{vonManteuffel:2014ixa, Peraro:2016wsq}.

The coefficient $r_i(\lambda, \tilde\lambda)$ in the finite remainder
are rational functions of the external kinematics. That is, they
belong to the field of fractions of the following polynomial quotient
ring \cite{DeLaurentis:2022otd},
\begin{equation}
  R_n = \mathbb{F}\big[|1⟩, [1|, \dots, |n⟩, [n|\big] \big/ \Big\langle \sum_i |i⟩[i| \Big\rangle \, ,
\end{equation}
where the big angle brackets denote an ideal, specifically the ideal
generated by the four momentum-conservation equations. The small angle
and square brackets denote spinors in the spinor-helicity notation,
$|i\rangle=\lambda_{i, \alpha}$ and $[i|=\tilde\lambda_{i,
    \dot\alpha}$. We refer to $R_n$ as the Lorentz covariant
  ring. Since the $r_i(\lambda, \tilde\lambda)$ are Lorentz invariant,
  they belong to a sub-ring of $R_n$, namely
\begin{equation}
  \mathcal{R}_n = \mathbb{F}\big[⟨1|2⟩, \dots, [n-1|n]\big] \big/ (\mathcal{J}_n + \mathcal{K}_n + \bar{\mathcal{K}}_n) \, ,
\end{equation}
where $\mathcal{J}_n$ denotes the Lorentz-invariant
momentum-conservation ideal, and $\mathcal{K}_n$ and
$\bar{\mathcal{K}}_n$ denote the ideals generated by holomorphic and
anti-holomorphic Schouten identities respectively.

\paragraph{Least common denominator}
In common-denominator form, the rational functions read,
\begin{equation}
  r_i(\lambda, \tilde\lambda) = \frac{\mathcal{N}_i(\lambda, \tilde\lambda)}{\prod_j
    \mathcal{D}_j^{\alpha_{ij}}(\lambda, \tilde\lambda)}
\end{equation}
where $\mathcal{N}_i(\lambda, \tilde\lambda)$ is a polynomial and the
denominator factors $\mathcal{D}_j(\lambda, \tilde\lambda)$ are
irreducible. If the powers $\alpha_{ij}\in \mathbb{Z}$ are as low as
possible, then the denominator is the least common denominator (LCD).

The poles $\mathcal{D}_j$ can be thought of as codimension-one
varieties $V(\big\langle \mathcal{D}_j \big\rangle)$ in $R_n$ (or
$\mathcal{R}_n$). At five points and up to two loops there are 35
possible poles, unchanged from one loop,
\begin{equation}
 \{\mathcal{D}_{\{1,\dots,35\}}\} = \bigcup_{\sigma \; \in \; \text{Aut}(R_5)} \sigma \circ \big\{ \langle 12 \rangle, \langle 1|2+3|1] \big\}
\end{equation}
where the automorphisms of $R_5$ are composed of permutations of the
legs $\{1,\dots,5\}$ and a possible swap of the left- and right-handed
representations of the Lorentz group, $\langle \rangle \leftrightarrow
[]$. The latter is related to parity.
Note that $\text{tr}_5=\text{tr}(\gamma_5p_1p_2p_3p_4)$ is not a
singularity of the pentagon-function coefficients in the finite
remainder. It is manifestly absent from the denominators of our
results.

\paragraph{Basis change}
The form of eq.~\ref{eq:LCD} is neither compact nor efficient for
reconstruction. In this form, we would have required around a quarter
million numerical probes. This is computationally prohibitive. In the
end, we managed to reconstruct the amplitude from around $35k$
numerical evaluations. To achieve this, we performed a basis change in
the vector space of pentagon coefficients, $\text{span}(r_i)$,
\begin{equation}\label{eq:basischangeschematic}
  \tilde r_{i} = O_{ij} \, r_{j\in \mathcal{B}} \, , \; O_{ij} \in \mathbb{Q} \, , \; \text{and} \; \mathcal{B} \text{ a basis of } \text{span}(r_i) \, .
\end{equation}
The aim is to minimise the denominator powers $\alpha_{ij}$ of the
basis functions. To achieve this, we require information regarding the
correlation of residues among the $r_i$'s at the different poles
$\mathcal{D}_k$. To obtain residues in $\mathbb{F}_p$ we perform an
univariate reconstruction and subsequent formal Laurent expansion,
\begin{equation}\label{eq:LCD}
  r_{i \in \mathcal{B}} = \sum_{m = 1}^{q_k = \text{max}_i(q_{ik})} \frac{e^k_{im}}{(t-t_{\mathcal{D}_k})^m} + \mathcal{O}((t-t_{\mathcal{D}_k})^0)
\end{equation}
If the denominator factors are not linear in $t$, this expansion can
be formulated as a $p(t)$-adic series \cite{Fontana:2023amt}. Through
Gaussian elimination on the residues $e^k_{im}$ on a set of different
slices we construct the following decomposition of the vector space,
\begin{equation}
  \text{span}(r_{i \in \mathcal{B}}) = \underbrace{\text{column}(\text{Res}(r_{i \in \mathcal{B}}, \mathcal{D}_k^m))}_{\text{functions with the singularity} \mathcal{D}_k^m} \;\;\; \oplus \, \underbrace{\text{null}(\text{Res}(r_{i \in \mathcal{B}}, \mathcal{D}_k^m))}_{\text{functions without the singularity} \mathcal{D}_k^m}
\end{equation}
We observe that the rational numbers appearing in the null spaces are
very simple and can be reconstructed with just a couple of
primes. Furthermore, the lifting $\mathbb{F}_p \rightarrow \mathbb{Q}$
of matrices in reduced row echelon form can be efficiently performed
by iterating on subsequent values of $p$ only those rows containing
entries deemed incorrect. A simple way to deem the correctness of a
reconstructed rational number is to re-scale it by a constant and
compare reconstructed results.


The matrix $O_{ij}$ is constructed by searching the space of
intersections of null spaces,
\begin{equation}
  \displaystyle O_{ij} = \bigcap_{k, m} \,
  \text{null}(\text{Res}(r_{i \in \mathcal{B}}, \mathcal{D}_k^m)) \, .
\end{equation}
We performed the basis change one sub-amplitude at a time for the
gluon channel, before merging them into all-plus, single-minus and MHV
vector spaces. The basis functions in LCD form after the basis change
are more than an order of magnitude simpler than the original ones.

\paragraph{Ansatz reconstruction}
Having determined a good set of functions to reconstruct, namely the
$\tilde r_i$ of eq.~\ref{eq:basischangeschematic}, we fit their
spinor-helicity Ansatz \cite{DeLaurentis:2019bjh, DeLaurentis:2022otd}
through $\mathbb{F}_p$ evaluations. The simplification stemming from
the basis change is compounded with that from a partial-fraction
decomposition (PFD). In our PFD we impose (and thus verify) that no
denominator contains more than a single letter of the equivalence
class with representative $\langle 1| 2+3| 1]$.

\paragraph{Little-group rescalings}
Having obtained a compact basis for the gluonic amplitude and inspired
by supersymmetry Ward identities, we note that candidate basis
functions for the quark amplitudes can be easily obtained by rescaling
the gluon basis functions by factors with suitable Little-group
weights \cite{DeLaurentis:2023izi}. Their validity is checked through
a far simpler computation than the full one.

\section{Conclusions and Outlook}
With the recent computation of double-virtual corrections to
$pp\rightarrow jjj$ \cite{Agarwal:2023suw, DeLaurentis:2023nss,
  DeLaurentis:2023izi} all two-loop five-point amplitudes with any
combination of partons and photons are known in full color in QCD
(excluding effects from quark masses, particularly contributions from
top quarks). This was possible thanks to three decades of progress in
perturbation theory since as the previous perturbative order for
$pp\rightarrow jjj$ was computed in 1993
\cite{Bern:1993mq,Bern:1994fz,Kunszt:1994nq}.

Our results, summarized in table~\ref{tab:VS-sizes}, demonstrate the
potential simplicity of scattering amplitudes even at second order in
perturbation theory thanks to the spinor-helicity formalism, partial
fraction decompositions, the manifestation of symmetries, and a
carefully chosen basis change in the vector space of rational
coefficients.

Future computations for multi-loop amplitudes with external or
internal masses, or higher multiplicities, will have to contend with a
steep increase in complexity not only due to the increased number of
scales but also due to more complicated alphabets and additional
redundancies in kinematic representations. In an upcoming publication
\cite{DeLaurentis:2024xxx} we demonstrate the efficacy of the
techniques described here, along with others, in achieving a drastic
simplification for the two-loop leading-color amplitudes for
$pp\rightarrow Wjj$.

\begin{table}[t]
\renewcommand{\arraystretch}{1.2}
\centering
\begin{tabular}{*{4}{C{15ex}}}
\toprule
Particle Helicities & Vector-space dimension & Generating set size & File size \\
\midrule
$g^+g^+g^+g^+g^+$ & 24 & 3 & 2 KB \\
$g^+g^+g^+g^+g^-$ & 440 & 33 & 24 KB \\
$g^+g^+g^+g^-g^-$ & 937 & 115 & 68 KB \\
\midrule
$u^+\bar{u}^-g^+g^+\,g^+$ & 424 & 91 & 45 KB \\
$u^+\bar{u}^-g^+g^+\,g^-$ & 844 & 449 & 200 KB \\
\midrule
$u^+\bar{u}^-d^+\bar{d}^-g^-$ & 435 & 124 & 56 KB \\
\bottomrule
\end{tabular}
\caption{
  \label{tab:VS-sizes} For each helicity configuration, this table
  shows the dimension of the vector space of rational functions, the
  number of functions in the generating set that spans the space upon
  closure under the symmetries of the vector of Little-group scalings
  of the given process, and the file size where the generating set is
  stored. This does not include the matrices of rational numbers
  needed to express the pentagon-function coefficients in terms of the
  basis of the vector space that they span. }
\end{table}

\paragraph{Acknowledgements}
I would like to thank H.~Ita and V.~Sotnikov for comments on this
manuscript and collaboration in the related computations and
publications.

\vspace{-1mm}

\setlength{\bibsep}{3pt}
\bibliography{main}

\end{document}